\documentclass[aps,prl,twocolumn,floatfix]{revtex4}
\pdfoutput=1
\usepackage{array}
\usepackage{amsmath}
\usepackage{graphicx}

\def\ket#1{|#1 \rangle}

\begin{document}

\title{Entangling quantum and classical states of light}

\author{Hyunseok Jeong$^{1}$, Alessandro Zavatta$^{2,3}$,  Minsu Kang$^{1}$, Seung-Woo Lee$^{1}$, Luca S. Costanzo$^{3}$, Samuele Grandi$^{2}$, Timothy C. Ralph$^{4}$ \&  Marco Bellini$^{2,3}$}

\affiliation{$^{1}$Center for Macroscopic Quantum Control, Department of Physics and Astronomy, Seoul National University, Seoul, 151-742, Korea\\
$^{2}$Istituto Nazionale di Ottica (INO-CNR), L.go E. Fermi 6, 50125 Florence, Italy\\
$^{3}$LENS and Department of Physics, University of Firenze, 50019 Sesto Fiorentino, Florence, Italy\\
$^{4}$Centre for Quantum Computation and Communication Technology, School of Mathematics and Physics, University of Queensland, Qld 4072, Australia.}

\begin{abstract}
Entanglement between quantum and classical objects is of special interest in the context of
fundamental studies of quantum mechanics and potential applications to quantum information
processing. In quantum optics, single photons are treated as light quanta while coherent states are
considered the most classical among all pure states. Recently, entanglement between a single photon and
a coherent state in a free-traveling field was identified to be a useful resource for
optical quantum information processing. However, it was pointed out to be extremely difficult to
generate such states since it requires a clean cross-Kerr nonlinear interaction. Here, we devise
and experimentally demonstrate a scheme to generate such hybrid entanglement by implementing a
coherent superposition of two distinct quantum operations.
The generated states clearly show entanglement between the two different types of states. Our work
opens a way to generate hybrid entanglement of a larger size and to develop efficient quantum
information processing using such a new type of qubits.
\end{abstract}

\maketitle

Quantum entanglement is of crucial importance for fundamental tests of quantum mechanics and
implementations of quantum information processing. The idea of entangling ``classical'' and
``quantum'' states is found in Schr\"odinger's famous cat paradox \cite{Schr}, where a microscopic
atom --as a quantum particle-- and a cat --as a classical object-- were assumed to be entangled to each
other. In quantum optics, coherent states are considered the most classical among all pure states \cite{Glauber}.
%
In many situations they can be treated semi-classically, {\it i.e.} as a classical light field with the addition of stochastic noise, and are typically most robust against decoherence
\cite{Zurek2003}. On the other hand, single photons are normally treated as discrete light quanta
containing the minimum quantized amount of energy available at a given frequency
and any attempt to describe them with an effective noise theory leads to negative probabilities.

Recently, entanglement between a single photon and a coherent state was identified to be
a very useful resource for optical quantum information processing, enabling one to perform nearly deterministic quantum teleportation and universal gate operations for quantum computation using linear optics \cite{LeeJeong2013}. This type of \emph{hybrid} entanglement, however, is difficult to generate in spite of its conceptual interest and potential usefulness. It is well known that a clean cross-Kerr type interaction between a single photon and a coherent state may generate such a state \cite{Gerry1999,Jeong2005}. However, many fundamental problems lie in the way of realizing a suitable interaction of this kind
\cite{Jeff2006,Jeff2007,JG2010}.
Therefore, an experimentally accessible scheme to replace the cross-Kerr nonlinearity and create entanglement between a single photon and a coherent state would be highly desirable.
In this article, we introduce such a scheme and use it to experimentally generate small-scale hybrid entanglement. We also outline extensions whereby our methods can be generalized to produce larger scale hybrid entanglement.

The hybrid entangled state considered in this article is
\begin{equation}
|\Psi(\alpha)\rangle= \frac{1}{\sqrt{2}}(|0\rangle|\alpha\rangle+|1\rangle|-\alpha\rangle)
\label{eq:hybrid}
\end{equation}
where $|0\rangle$ and  $|1\rangle$ are the vacuum and single photon state, respectively, and
$|\pm\alpha\rangle$ are coherent states of amplitudes $\pm\alpha$. Due to the classical properties
of coherent states, state (\ref{eq:hybrid}) can be considered as the closest optical implementation of
Schr\"odinger's cat Gedanken-experiment, as it manifests the entanglement between quantum and
classical states, particularly when $\alpha$ is sufficiently large. It is straightforward to prove
that its degree as a macroscopic superposition \cite{LeeJeong2011} increases as $\alpha$ becomes
larger, and it actually shows the maximum possible value, {\it i.e.}, equal to its average photon
number. This is also true for some highly nonclassical macroscopic states \cite{LeeJeong2011} such
as a coherent-state superposition, $|\alpha\rangle+|-\alpha\rangle$ \cite{Ourj2007}, and a photon number entanglement called the ``NOON state'' \cite{Afek2010}.
On the other hand, states generated by applying the displacement operation on one part of
single-photon entanglement \cite{Seka2012,Bruno2013,Lvovsky2013} do not show such properties
because a stringent degree of macroscopic superposition for a bosonic state is shown to be
invariant under the displacement operation \cite{LeeJeong2011}.

In order to generate hybrid entanglement, our first observation is that a photon-added coherent
state, the result of the (in principle, multiple) application of the photon creation operator onto a coherent state, can be a good approximation of a coherent state of a larger amplitude as
\begin{equation}
{\cal N}^{-\frac{1}{2}}
(\hat{a}^\dagger)^n|\alpha\rangle\approx|g\alpha\rangle
\label{eq:c-approx}
\end{equation}
where ${\cal N}
=n!L_n(-|\alpha^2|)$, with the Laguerre polynomial of nth order $L_n$,
\begin{equation}
g=\frac{1}{2}+\sqrt{\frac{1}{4}+\frac{n}{|\alpha|^2}}~,
\end{equation}
and this approximation holds when $\alpha$ is not too small. The fidelity between
state~(\ref{eq:c-approx}) and an ideal coherent state $|g\alpha\rangle$ is
\begin{equation}
{\cal F}=\mathcal{N}^{-1}
g^{2n}|\alpha|^{2n}e^{-|\alpha|^2(g-1)^2}.
\end{equation}
Here we shall focus on the particular case of $n=1$, {\it i.e.}, the single-photon addition to a coherent state. In this case, for example, ${\cal F}\approx0.98$ for $\alpha=2$ and $g\alpha= 2.414$, and ${\cal F}\approx0.998$ for $\alpha=4$ and $g\alpha= 4.236$. In fact, conditional single-photon addition can be experimentally implemented in the signal mode of a parametric down-converter upon detection of an idler photon, and this approach has already been used for fundamental proofs of quantum mechanics \cite{Zavatta2004,Parigi2007,KimReview2008}.

\begin{figure*}[t]
\includegraphics[width=0.9\linewidth]{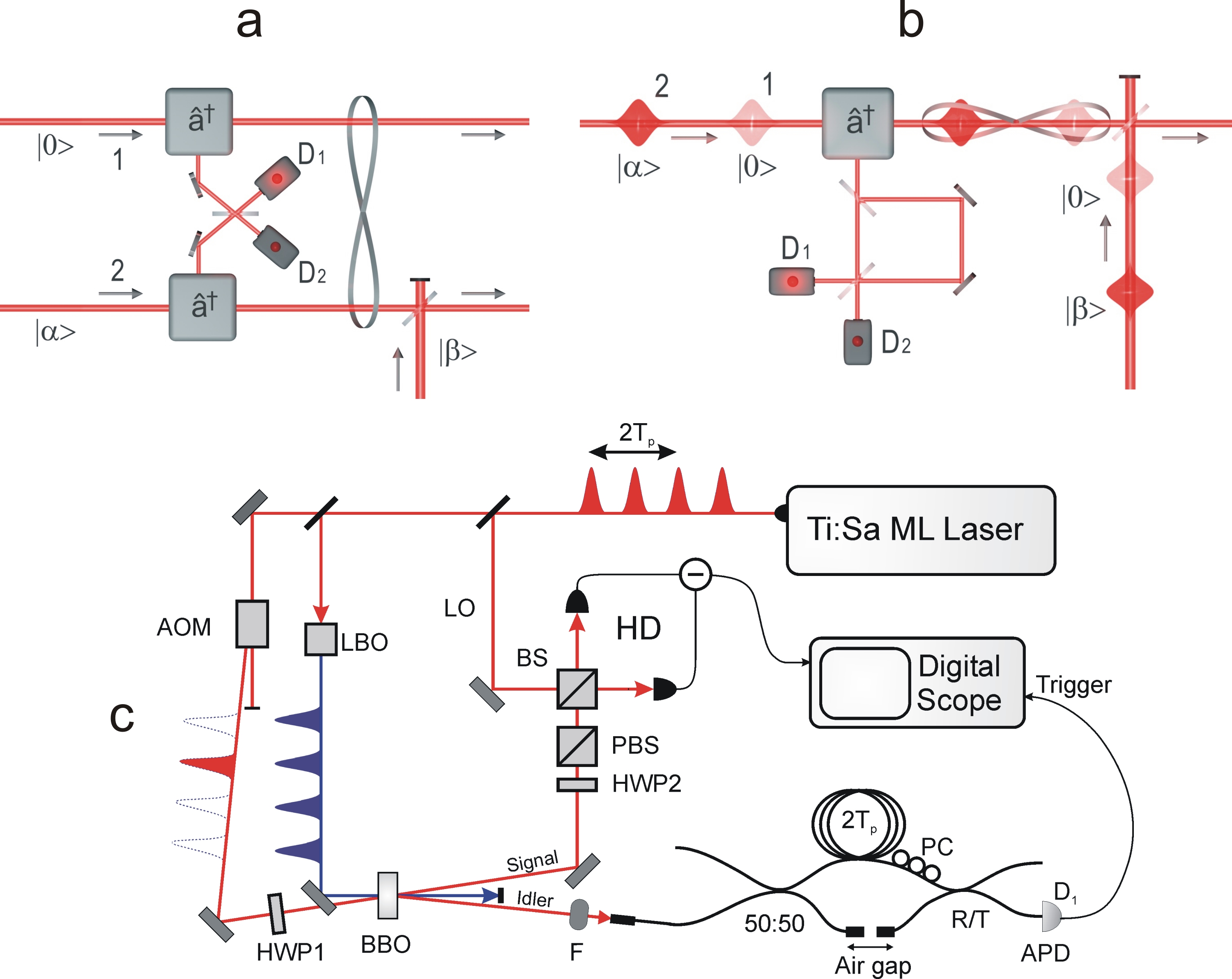}
\caption{a. Conceptual schematic for generating small-scale hybrid entanglement by superposing two
photon-addition operations on different spatial modes. The phase-space displacement operation on mode 2 is then obtained by mixing with an additional intense coherent field ($\ket{\beta}$) on a low-reflectivity beam-splitter. D$_1$ and D$_2$ are two single-photon detectors. b. Temporal-mode version of the scheme, as used in the experiment. Here, an unbalanced Mach-Zehnder interferometer makes the 'clicks' heralding the photon addition onto the two temporal modes, 1 and 2, indistinguishable. This results in a superposition of photon-addition operations on the two modes, containing vacuum ($\ket{0}$) and a coherent state ($\ket{\alpha}$), respectively. The state displacement is again obtained by mixing with an additional intense coherent state ($\ket{\beta}$), only synchronized with the second temporal mode. c. Actual experimental scheme. $T_P$ is the delay between successive pulses from the mode-locked laser; PC is a fiber polarization controller; R/T indicates a variable-ratio fiber beam-splitter; D$_1$ is a fiber coupled avalanche photodiode (APD); HWP1 and HWP2 are half-wave plates, and PBS is a polarizing beam-splitter used for the displacement operation. The locking mechanism of the fiber Mach-Zehnder interferometer is not shown for the sake of clarity. (See Methods for further details).}
    \label{keyidea}
\end{figure*}

The next point of our scheme is to implement a quantum superposition of distinct operations as
\begin{equation}
r{\hat a}_1^\dagger + t  {\hat a}_2^\dagger \label{eq:so}
\end{equation}
where $r=\sqrt{1-t^2}$ and 1 and 2 are two distinct modes. These types of ``superposed'' operations can be implemented by using two photon-addition units, a beam-splitter of reflectivity $r$ and two single-photon detectors, as shown in Fig.~\ref{keyidea}a. The key idea is to mix the heralding idler photons of the two processes on the beam-splitter to erase the information as to which of ${\hat a}^\dagger_1$ or ${\hat a}^\dagger_2$ occurred, while one of those two events definitely happened. Similar schemes for superposing different quantum operations, but acting on the same mode, were recently successfully applied for the direct experimental test of quantum commutation rules \cite{Kim2008,Zavatta2009}.

If the operation (\ref{eq:so}) is applied to an initial state of $|0\rangle_1\otimes|\alpha_i\rangle_2$, photon addition can either take place in the first mode, producing a single-photon Fock state $\ket{1}_1$ while leaving the coherent state $\ket{\alpha_i}_2$ in the second mode, or in the second mode, leaving the vacuum in the first mode and producing a photon-added coherent state in the second.
By setting $t=\{\alpha^2+2\}^{-1/2}$, one can balance the probabilities of these two events and the resulting state is
\begin{equation}
\begin{aligned}
&\frac{1}{\sqrt{2}}\Big(|1\rangle|\alpha_i\rangle+|0\rangle\frac{\hat{a}^\dagger|\alpha_i\rangle}{\sqrt{|\alpha_i|^2+1}}\Big)\\
&\approx\frac{1}{\sqrt{2}}\Big(|1\rangle|\alpha_i\rangle+|0\rangle|g\alpha_i\rangle\Big),
\end{aligned}
\label{eq:ssp}
\end{equation}
which clearly presents a desired form of hybrid entanglement between a single-photon quantum and a coherent-state field.

\begin{figure*}[t]
\includegraphics[width=0.85\linewidth]{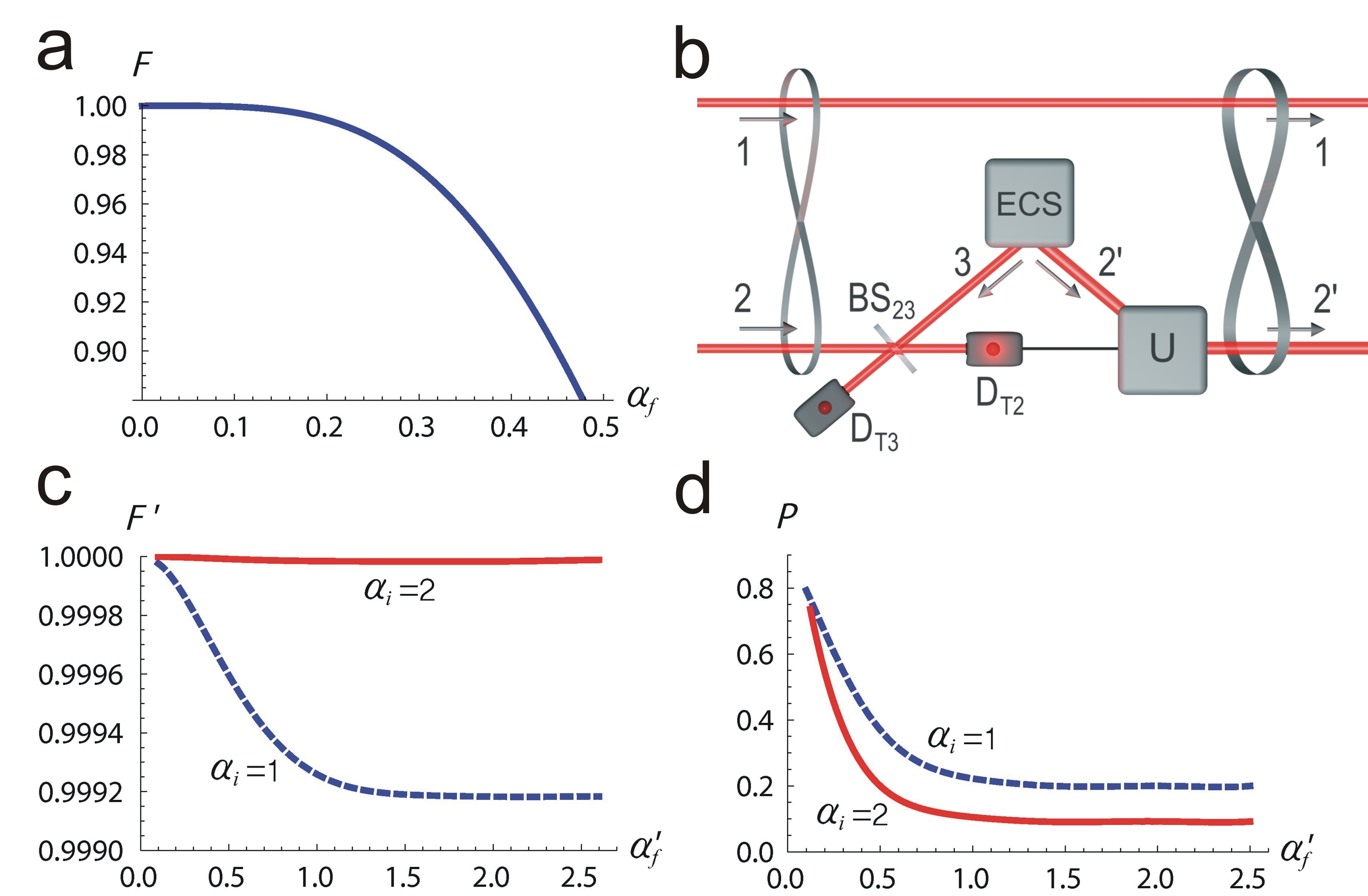}
 \caption{a. Fidelity $F$ for small-size hybrid entanglement against the output amplitude $\alpha_f$ optimized over the initial amplitude $\alpha_i$. b. Conceptual scheme for the generation of larger-size hybrid entanglement by tele-amplifying the coherent-state part (mode 2) of small-size hybrid entanglement. c. Fidelity $ F'$ for larger-size hybrid entanglement against the amplified amplitude $\alpha_f'$ for initial amplitudes $\alpha_i=1$ (dashed) and $\alpha_i=2$ (solid). Nearly perfect fidelities are obtained using the tele-amplification method. d. Success probability $ P$ of the tele-amplification process to obtain larger-size hybrid entanglement against the output amplitude  $\alpha_f'$ for initial amplitudes $\alpha_i=1$ (dashed) and $\alpha_i=2$ (solid).}\label{fig:fidelity}
\end{figure*}

In order to make a small-scale hybrid entangled state in the symmetric form of Eq.~(\ref{eq:hybrid}) for quantum
information processing \cite{LeeJeong2013}, the displacement operation $D(-\frac{\alpha_i+g\alpha_i}{2})$ may be finally applied to the coherent-state part in the second mode. The state then becomes
\begin{equation}
|\Psi_S\rangle_{12}\approx\frac{1}{\sqrt{2}}(|0\rangle|\alpha_f\rangle+|1\rangle|-\alpha_f\rangle)
\label{eq:symstate}
\end{equation}
where $\alpha_f=(g\alpha_i-\alpha_i)/2$.
It should  be noted that the local displacement operation here changes {\it neither} the degree of entanglement \cite{Peres-NPT, Horodecki-NPT, Lee-NPT} {\it nor} that of macroscopic quantumness \cite{LeeJeong2011}.
The fidelity between state $|\Psi_S\rangle$ and the ideal hybrid one is
\begin{equation}
\begin{aligned}
F&=|\langle\Psi_S |\Psi(\alpha_f)\rangle|^2\\
&= \frac{1}{4}\Big( 1+\frac{(g\alpha_i)^2 e^{-4\alpha_f^2}}{1+\alpha_i^2} +\frac{2g\alpha_i
e^{-2\alpha_f^2}}{\sqrt{1+\alpha_i^2}} \Big)
\end{aligned}
\end{equation}
and can be made high for properly chosen values of $\alpha_i$ and $\alpha_f$, as presented in 
Fig.~\ref{fig:fidelity}a.
For example, the fidelity is expected to be as high as $F\approx0.991$ with $\alpha_f\approx0.21$ when the initial amplitude was $\alpha_i=2$. A smaller initial amplitude $\alpha_i$ may increase the effective amplitude $\alpha_f$ of the final hybrid state, but at the price of a lower fidelity to the ideal one ($F\approx0.946$ with $\alpha_f\approx0.31$ when $\alpha_i=1$).
Very recently, a different scheme to generate the hybrid state (\ref{eq:hybrid}) was suggested by Andersen and Neergaard-Nielsen \cite{Andersen2013}, but their approach requires a coherent-state superposition and a Hadamard transform for $|0\rangle$ and $|1\rangle$ in addition to the resources used in our experiment.

An amplitude $\alpha_f\approx 0.21$ for a hybrid pair is still small if compared to the best value for quantum information processing of $\approx1$ \cite{LeeJeong2013}. In order to obtain larger-size hybrid entanglement with a high fidelity, one may increase the amplitude $\alpha_f$ of the hybrid entangled state by using the tele-amplification scheme \cite{ralph-teleamp, teleamp2013}.

As shown in Fig.~\ref{fig:fidelity}b, the coherent-state part (mode 2) of the hybrid pair $|\Psi(\alpha_f)\rangle$ can be teleported using an entangled coherent state as
the teleportation channel
\begin{equation}
|{\rm ECS}\rangle_{ 3 2'}={\cal N}'(|\alpha_f\rangle_3|\alpha_f'\rangle_{2'}-|-\alpha_f\rangle_3|-\alpha_f'
\rangle_{2'})
\end{equation}
where ${\cal N'}=[2(1-e^{-2\alpha_f^2-2\alpha_f'^2})]^{-1/2}$ and $\alpha_f'>\alpha_f$. The
tele-amplification process is successful if one of the detectors, $\rm D_{T2}$ or $\rm D_{T3}$, registers
a single photon \cite{Jeong2001,Enk2001,teleamp2013}. The last requirement is that, if $\rm D_{T2}$
clicks, a $\pi$ phase shift (U) has to be performed to complete the tele-amplification process
\cite{Jeong2001,Enk2001}.

Remarkably, the tele-amplification can increase the fidelity of the obtained tele-amplified hybrid entangled state with the ideal one $|\Psi(\alpha_f')\rangle$ up to $F'>0.9999$  ($F'>0.999$) for the initial amplitude $\alpha_i=2$ ($\alpha_i=1$) regardless of the value of $\alpha_f$, as presented in Fig.~\ref{fig:fidelity}c.
This is because the single-photon  measurement distills off the errors originated from the approximation
$\hat{a}^\dagger |\alpha_i\rangle\propto |g\alpha_i\rangle$ in $|\Psi_S\rangle$ by joint projection
on the Bell basis with the prepared entangled coherent channel $|{\rm ECS}\rangle$ \cite{teleamp2013,Jeong2002QIC}. The fidelity can be made even higher by choosing a larger value of $\alpha_i$ for the initial coherent state at the price of a lower success probability (see Fig.~\ref{fig:fidelity}d and Methods).

\bigskip

We now discuss our experimental demonstration of small-scale hybrid entanglement. Differently from the schematic picture of Fig.\ref{keyidea}a, our actual experimental setup makes use
of two distinct traveling temporal modes (instead of spatial ones) for encoding the entangled state,
and is based on a scheme first developed for the remote generation of coherently-delocalized single
photons~\cite{Zavatta06,Dangelo06}. This approach allows us to use a single photon-addition device and a single homodyne detector, but in a time-multiplexed fashion, to conditionally generate and analyze the two-mode state (see Methods).

As shown schematically in Fig.\ref{keyidea}b, we pass the idler photons produced in a photon-addition device based on parametric down-conversion \cite{Zavatta2004} through an unbalanced Mach-Zehnder interferometer. If an idler photon is detected at one of the exits of the interferometer and it is not possible, even in principle, to discern whether it passed through the long or the short interferometer arm, then the superposition $r{\hat a}_{t_1}^\dagger + t  {\hat a}_{t_2}^\dagger$ of two single-photon additions onto two signal temporal modes, denoted by $t_1$ and $t_2$, and separated by the interferometer delay $(t_2-t_1)$, is effectively heralded. The analogy with the general quantum operation (\ref{eq:so}) is complete if we note that the varying $r$ and $t$ parameters can be simply implemented by adjusting the transmission and the relative phase of the two interferometer arms.
The temporal-mode version of the entangled state (\ref{eq:ssp}) is obtained by seeding the signal mode of the parametric down-converter with the state $\ket{0}_{t1}\otimes\ket{\alpha_i}_{t2}$, containing vacuum in the first temporal mode and a coherent state in the second.
Time-domain homodyne detection then measures pairs of quadrature values, corresponding to the two distinct temporal modes of the state, for each heralding event (see Methods). A full quantum tomographic reconstruction of the two-mode density matrix of the states is finally performed, based on a purposely-developed, iterative, max-likelihood algorithm \cite{lvovsky04,hradil06}.
\begin{figure*}[t]
\includegraphics[width=1\linewidth]{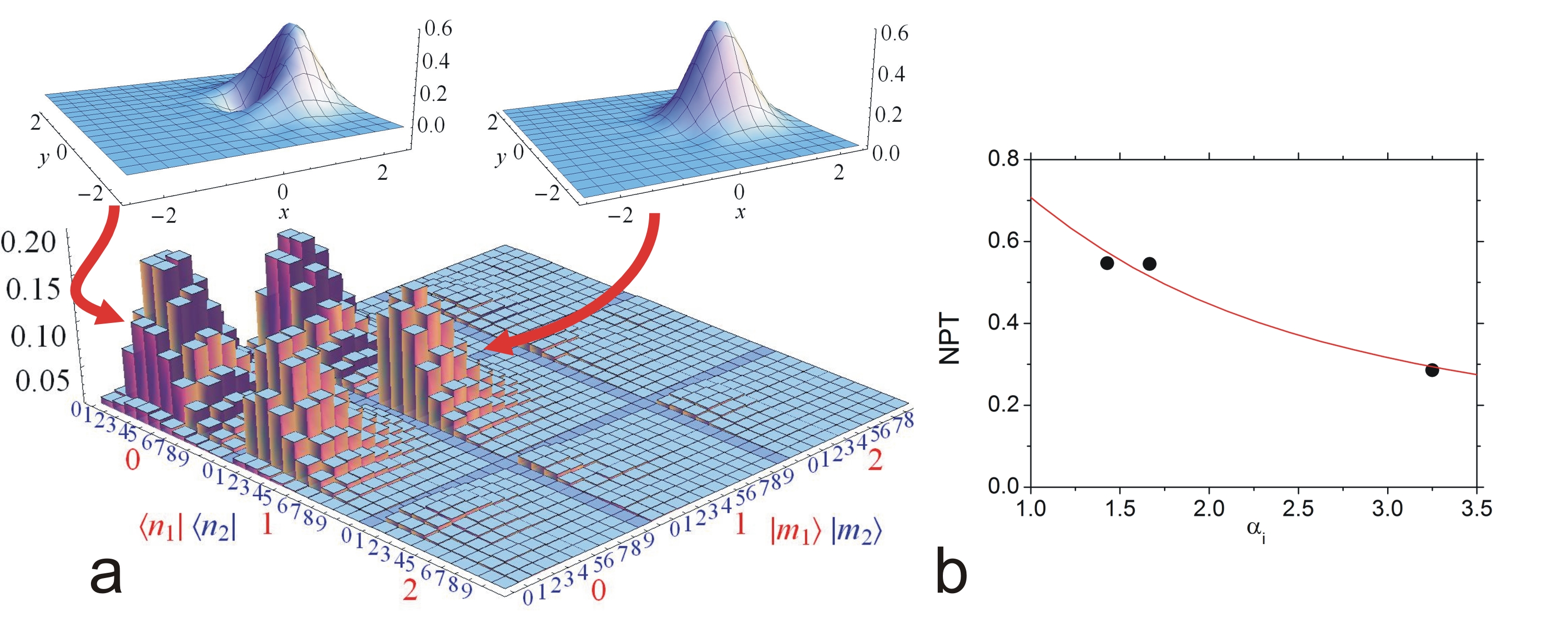}
\caption{a. Reconstructed density matrix for the experimentally generated hybrid entangled state of Eq.\ref{eq:ssp} with $\alpha_i \approx 1.4$. In this case we use $3\times 3$ blocks of $10 \times 10$ matrix elements, and we correct for a detection efficiency of $61\%$. The two insets show the Wigner functions for the state in the second mode when vacuum or a single photon are observed in the first. b. Plot of the negativity of the partial transpose (NPT) as a function of the initial coherent state amplitude $\alpha_i$. Solid curve: calculated NPT for the expected state (unit efficiency); data points: NPT derived from experimentally reconstructed density matrices, corrected for detection efficiency. Error bars of statistical origin, related to the number of homodyne data acquisitions, are of the order of 2$\%$ and within the size of the data points.}
    \label{fig:densmat}
\end{figure*}

Fig.\ref{fig:densmat}a shows the reconstructed density matrix for a typical balanced state of the kind of Eq.(\ref{eq:ssp}). Roughly speaking, the blocks on the diagonal tell us that the second mode contains a photon-added coherent state (ideally lacking the vacuum contribution) when the first mode is empty, while it contains a coherent state when a single photon excites the first mode. This is also shown in the two insets of Fig.\ref{fig:densmat}a, where we plot the Wigner functions of the states the second mode is projected to when we impose the first mode to be in a vacuum or single-photon state, respectively.
The non-vanishing off-diagonal blocks in the density matrix demonstrate that we are not dealing with a statistical mixture of these two situations, but with a coherent superposition of the two, that is an entangled state like the one described by (\ref{eq:ssp}). The negativity of the partial transpose (NPT), a measure of the degree of entanglement of the state \cite{Lee-NPT}, calculated from the experimentally reconstructed density matrix in this case ($\alpha_i \approx 1.4$) is NPT=0.55, in fair agreement with the theoretical expectations, considering the non-unit state preparation efficiency. Similar results were obtained for values of $\alpha_i$ ranging from 1.4 to 3.25 (as shown in Fig.\ref{fig:densmat}b), demonstrating that we are indeed able to experimentally generate hybrid entanglement between a quantum (discrete) single-photon qubit state, and a classical (continuous) one, made of two coherent states of different amplitudes, $\ket{\alpha_i}$ and $\ket{g \alpha_i}$.

\begin{figure}[t]
\includegraphics[width=1\linewidth]{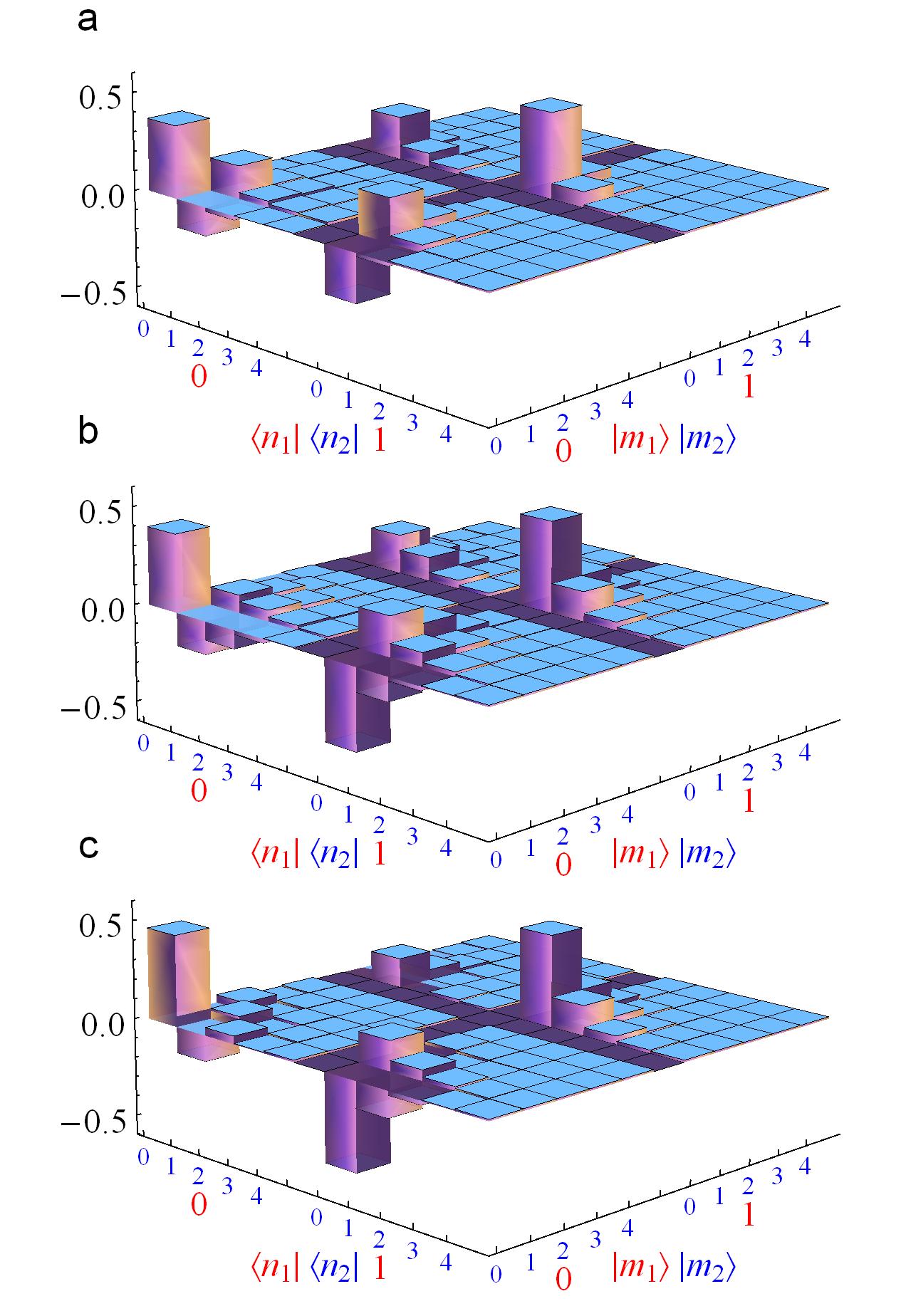}
\caption{a. Experimentally reconstructed density matrix for a symmetric, small-scale, hybrid entangled state, as obtained after phase-space coherent displacement. A detection efficiency of $63\%$ has been corrected for in the reconstruction. Because of the small effective amplitude ($\alpha_f\approx 0.31$), the number of reconstructed elements in the Fock space has been decreased correspondingly. b. Calculated density matrix for the expected $|\Psi_S\rangle_{12}$ state (with unit efficiency). c. Calculated density matrix for the ideal hybrid state $|\Psi(\alpha_f)\rangle$ with $\alpha_f=0.31$.}
    \label{fig:densmatdisp}
\end{figure}

To produce the symmetric hybrid state (\ref{eq:symstate}) we experimentally implemented the displacement operation on the second temporal mode by combining the signal beam with properly synchronized intense coherent pulses (indicated as $\ket{\beta}$ in Fig.\ref{keyidea}b on a low-reflectivity beam-splitter in a polarization scheme (see Methods) \cite{Lvovsky2013}.
The reconstructed density matrix for such a symmetric hybrid state, with $\alpha_i \approx 1$ and an effective final amplitude $\alpha_f\approx 0.31$, is shown in Fig.\ref{fig:densmatdisp}a together with the calculated ones corresponding to the expected (b) and ideal cases (c).
Despite the difficulties in experimentally providing the exact amount of phase-space displacement and in stabilizing the phase of such a low-amplitude state, we still find a substantial degree of entanglement (NPT=0.45) in the reconstructed state, which has a fidelity $F=0.92$ to the model state expected under the same experimental conditions.


\bigskip

In conclusion, we have devised a scheme to generate optical hybrid entanglement by quantum-mechanically superposing non-Gaussian operations on distinct modes, and we experimentally performed its small-scale demonstration. Our proposal may enable one to obtain larger-scale hybrid entanglement without using cross-Kerr nonlinearity and with currently available resources. In fact, coherent-state superpositions up to $|\alpha|^2 \approx 2.6$ have been experimentally realized \cite{Ourj2007,Sasaki2008,Gerrits2010}. This means that an unbalanced entangled coherent state $|{\rm
ECS}\rangle\propto |\alpha_f\rangle|\alpha_f'\rangle-|-\alpha_f\rangle|-\alpha_f'\rangle$ with $\alpha_f\approx0.31$ and $\alpha_f'\approx1.58$ may be obtained using an unbalanced beam splitter.
Therefore, as the next step to produce larger-size hybrid entanglement, the tele-amplification of hybrid entanglement from $\alpha_f\sim 0.3$ to $\alpha_f'\sim 1.5$ may be a reasonable task using current technology.

As the first experimental demonstration of entanglement between a single photon and a coherent-state field,
our work paves the way to experimentally explore quantum-classical entanglement. Furthermore, it constitutes an important step forward to new forms of efficient, all-optical, quantum information processing. Encoding qubits in such a hybrid form, as it effectively combines the intrinsic advantages of each part, enables one to realize quantum teleportation and a universal set of gate operations for quantum computing in a nearly deterministic manner using linear optics elements \cite{LeeJeong2013}.

\newpage

\subsection{Methods}

\subsubsection{Experimental setup}

We use the train of $1.5$-ps pulses emitted at $785$~nm from a mode-locked Ti:Sapphire laser at a repetition rate of $80$~MHz as the main light source for the experiment (see Fig.\ref{keyidea}c). Most of the emission is frequency doubled in a lithium triborate (LBO) crystal to become the pump for degenerate non-collinear type-I parametric down-conversion (PDC) in a $\beta$-barium borate (BBO) crystal.

Idler photons are spectrally filtered with a pair of etalon cavities (F), sent to an unbalanced Mach-Zehnder interferometer based on single-mode fibers and variable-ratio fiber couplers, and finally detected at one of the two exit ports of the interferometer by an avalanche photodiode D$_1$. The optical path difference between the two interferometer arms is made equal to twice the distance between successive pulses in the laser train, corresponding to a delay of about 24 ns. A fine adjustment of the delay and an accurate control of the relative phase between the two paths is then obtained by means of a short free-space propagation and piezoelectric transducers.

A small part of the emission from the mode-locked laser is passed through an acousto-optic modulator (AOM) working as a pulse-picker to select one in every eighth pulse of the train. This train of attenuated laser pulses at a lower repetition rate ($10$~MHz) is injected in the signal spatial mode of the BBO crystal to provide the seed coherent state $\ket{\alpha_i}$ in the temporal mode 2 and the vacuum state $\ket{0}$ in mode 1 for photon addition.

The idler detector D$_1$ is gated by the AOM modulation in such a way that, when the long interferometer arm is blocked, its clicks herald the addition of a single photon to one of the coherent states in the $10$~MHz seed train. In the same timing conditions, but with only the short arm blocked instead, a D$_1$ click heralds the production of a single photon from the vacuum state in the temporal mode arriving 24 ns earlier. Therefore, when both interferometer arms are open, a click in the idler detector cannot distinguish between the following two situations: a) a spontaneous down-conversion event was produced by the $n^{th}$ pump pulse, the idler photon traveled the long arm, and therefore a single-photon Fock state is present in mode 1, while a coherent state is in mode 2; b) a stimulated down-conversion event was produced by the $(n+2)^{th}$ pump pulse, the idler photon traveled the short arm, and therefore the vacuum state is present in mode 1, while a photon-added coherent state is in mode 2.

Because of this indistinguishability, a coherent superposition of the two final states in the two different temporal modes is obtained. Since the photon addition to a coherent state relies on stimulated rather than spontaneous emission, the weights of the two terms of the entangled state are not equal and have to be balanced by adjusting the splitting ratio of the second fiber coupler according to the coherent state amplitude $|\alpha_i|$. This is done experimentally by equalizing the idler count rates in D$_1$ when the two interferometer arms are alternatively blocked.

Note that, differently from the schematic views of Figs.\ref{keyidea}a and b, two single-photon detectors are not necessary in either possible version of the experiment. Because of the very low gain of the parametric crystal used for heralded photon addition, the probability of the two detectors clicking together (for a two-photon addition event or for the simultaneous addition of a photon to both modes) is completely negligible in our case.

Differently from the states described by Eqs.\ref{eq:ssp} and \ref{eq:symstate}, in this experiment we can also remotely introduce an arbitrary relative phase between the two terms of the signal entangled state by acting on the Mach-Zehnder interferometer in the idler path. This phase is actively controlled and locked to 9 values in the [0, $\pi$] interval by injecting counter-propagating laser pulses into the unused exit port of the interferometer while monitoring the interference signal at the spare input port.

In order to perform temporal-mode-selective coherent displacement of the state, we also make use of the polarization degree of freedom of the light pulses in the experiment. The correct amount of phase-space displacement is obtained by combining the vertically-polarized signal field with an intense, co-propagating, horizontally-polarized, version of the coherent seed pulse train at $10$~MHz on a slightly rotated half-wave plate. Being synchronized with the second temporal mode only, the displacement coherent pulses do not affect the single-photon part of the state in the first temporal mode.

\subsubsection{Homodyne detection and state tomography}

After the PDC crystal, the signal beam propagates in free space before being mixed at a 50:50 beam-splitter (BS) with intense, coherent, local oscillator (LO) pulses from the mode-locked laser for high-frequency time-domain balanced homodyne detection (HD) \cite{zavatta02:josab}. The global relative phase between the LO pulse train (at $80$~MHz repetition rate) and the states to be measured is actively locked to 7$\div$13 (depending on the coherent state amplitude) different values in the [0, $\pi$] interval by monitoring the dc level from the homodyne receiver. Triggered by a D$_1$ click, two different, phase-locked, pulses in the LO train are used to perform quadrature measurements on the two temporal modes of interest (corresponding to the $n^{th}$ and the $(n+2)^{th}$ pump pulses), by means of a digital oscilloscope. Strictly speaking, each data pair corresponds to a measurement of the same-phase quadrature for the two different modes of the entangled state. However, since we are able to produce entangled states with an arbitrary relative phase, it can be shown that this is equivalent to analyzing different-phase quadratures of states with a fixed relative phase, like those described in Eqs.\ref{eq:ssp} and \ref{eq:symstate}. About $6 \times 10^5$ quadrature pairs are acquired to perform the full tomography of each state. The dimensions of the reconstructed density matrices in the Fock basis are adjusted to the size of the investigated states. Therefore, three terms in the Fock expansion are normally used for the first mode (ideally containing just vacuum and single-photon components), whereas up to 25 may be necessary for the second mode, containing the coherent and the photon-added coherent states.

\subsubsection{Tele-amplification}

Let us assume that $|\Psi_S\rangle_{12}$ was generated by the process described in Fig.~\ref{keyidea}a
and tele-amplified using the entangled coherent channel $|\rm ECS\rangle$, as shown in Fig.~\ref{fig:fidelity}b.
After a 50:50 beam splitter, ${\rm BS}_{23}$, is applied to a part (mode $3$) of state $|\rm ECS\rangle_{32'}$ and mode 2, the total state becomes ${\hat B_{23}}|\Psi_S\rangle_{12}|{\rm
ECS}\rangle_{32'}$ where ${\hat B_{23}}$ represents the action of a 50:50 beam splitter on modes 2
and 3. Detectors ${\rm D}_{T2}$ and ${\rm D}_{T3}$ are then set to perform a type of Bell-state
measurement that in principle identifies all four Bell states \cite{Jeong2001,Jeong2002QIC},
$\ket{\alpha_f}\ket{\alpha_f}\pm\ket{-\alpha_f}\ket{-\alpha_f}$ and
$\ket{\alpha_f}\ket{-\alpha_f}\pm\ket{-\alpha_f}\ket{\alpha_f}$ (without the normalizations). In
our study, assuming availability of two single photon detectors, we employ $(1,0)$ and $(0,1)$ as
success events, which distinguishes two of the Bell states. One of the successful events leads to
\begin{equation}
|\Psi'_S\rangle_{12'}={\cal N}~_2\langle1|_3\langle0|{\hat B_{23}}|\Psi_S\rangle_{12}|{\rm
ECS}\rangle_{32'}
\end{equation}
and the other success case gives exactly the same result
\begin{equation}
|\Psi'_S\rangle_{12'}= {\cal N} U_{2'}(\pi)~_2\langle0|_3\langle1|{\hat
B_{23}}|\Psi_S\rangle_{12}|{\rm ECS}\rangle_{32'}
\end{equation}
where $U_{2'}(\pi)$ is a $\pi$ phase shift on mode $2'$. An explicit form of the result is
\begin{equation}
\begin{aligned}
&|\Psi_S'\rangle_{12'}={\cal N}\Big\{(g\alpha_i-\alpha_i)e^{-2\alpha_f^2}\ket{0}\ket{\alpha_f'}
\\&+\Big[(g\alpha_i e^{-4\alpha_f^2}-\alpha_i)\ket{0}
-\frac{\sqrt{1+\alpha_i^2}}{2{\cal N}_o^2} \ket{1}\Big]\ket{-\alpha_f'}\Big\}
\label{eq:teamp-state}
\end{aligned}
\end{equation}
where ${\cal N}_o=1/\sqrt{2(1-e^{-4\alpha_f^2})}$ and ${\cal N}$ is the normalization factor. We
then calculate its fidelity $F'$ to the ideal hybrid state $|\Psi(\alpha_f')\rangle$ as
\begin{equation}
\begin{aligned}
F'&=\frac{{\cal N}^2}{2}\Big\{(g\alpha_i-\alpha_i)e^{-2\alpha_f^2}
\\&+(g\alpha_ie^{-4\alpha_f^2}-\alpha_i)e^{-2{\alpha'_f}^2}+\frac{\sqrt{1+\alpha_i^2}}{2{\cal N}_o^2}\Big\}^2.
\label{ff}
\end{aligned}
\end{equation}

The success probability for the tele-amplification process can be obtained by summing the
probabilities in which only one of the detectors, either $\rm D_{T2}$ or $\rm D_{T3}$,  registers a photon. The probability in which only $\rm D_{T2}$
registers a photon is
\begin{equation}
P_{(1,0)}
=|_2\langle 1|_3\langle 0|{\hat B_{23}}|\Psi_S\rangle_{12}|{\rm ECS}\rangle_{2'3}|^2.
\end{equation}
The probability $P_{(0,1)}$ for the the case with  ``${\rm D}_{T3}$ click'' can be obtained in the
same manner so that $P=P_{(1,0)}+P_{(0,1)}$ in Eq.~(\ref{eq:sp}) is obtained as
\begin{equation}\label{eq:sp}
\begin{aligned}
P
&=\frac{{\cal N'}^2e^{-2\alpha_f^2}}{2(\alpha_i^2+1)}\Big\{1+(1-\frac{g\alpha_i^2-\alpha_i^2}{2})^2\\
&+4(\alpha_i^2+1)\alpha_f^2-2(1-\frac{g\alpha_i^2-\alpha_i^2}{2})e^{-2\alpha_f'^2}\Big\}.
\end{aligned}
\end{equation}
 In Fig.~\ref{fig:fidelity}c, we plot the numerically optimized success probability for Eq.~(\ref{eq:sp}) against $\alpha_f'$
for two chosen values of the initial amplitude $\alpha_i$.

\newpage


\bigskip
\bigskip

\section{Acknowledgments}A.Z., L.C., S.G., and M.B. acknowledge the support of Ente Cassa di Risparmio di Firenze, of the EU under the ERA-NET CHIST-ERA project QSCALE, and of the MIUR under the contract FIRB RBFR10M3SB. H.J., M.K., and S.-W. L were supported by the National Research Foundation of Korea (NRF) grant funded by the Korea government (MSIP) (No. 2010-0018295). T.C.R. acknowledges the support of the Australian Research Council Centre of Excellence for Quantum Computation and Communication Technology (Project number CE110001027).


\section{Correspondence} Correspondence and requests for materials should be addressed to M.B. (email: bellini@ino.it) or H.J (email: h.jeong37@gmail.com).

\end{document}